\documentclass[aps,preprint,groupedaddress]{revtex4-1}

\usepackage{amsmath} % AMS Math Package
\usepackage{amssymb}	% Math symbols such as \mathbb
\usepackage{graphicx} % Allows for eps images

\renewcommand{\v}[1]{\ensuremath{\mathbf{#1}}} % for vectors
\newcommand{\uv}[1]{\ensuremath{\mathbf{\hat{#1}}}} % for unit vector
\renewcommand{\d}[2]{\frac{d #1}{d #2}} % for derivatives
\newcommand{\pd}[2]{\frac{\partial #1}{\partial #2}} 
\newcommand{\pdd}[2]{\frac{\partial^2 #1}{\partial #2^2}} 

\begin{document}

\newpage

\title{Traveling Wave Model for Frequency Comb Generation in Single Section Quantum Well Diode Lasers}

\author{Mark Dong}
\email{markdong@umich.edu}
\affiliation{Department of Electrical Engineering and Computer Science, University of Michigan, \\1301 Beal Avenue, Ann Arbor, 48109-2122}
\affiliation{Department of Physics, University of Michigan, 450 Church Street, Ann Arbor, MI 48109-1040}

\author{Niall M. Mangan}
\email{niallmm@gmail.com}
\affiliation{Department of Applied Mathematics, University of Washington\\ Lewis Hall 202, Box 353925, Seattle, WA 98195-3925 }

\author{J. Nathan Kutz}
 \email{kutz@uw.edu}
\affiliation{Department of Applied Mathematics, University of Washington\\ Lewis Hall 202, Box 353925, Seattle, WA 98195-3925 }

\author{Steven T. Cundiff}
 \email{cundiff@umich.edu}
\affiliation{Department of Physics, University of Michigan, 450 Church Street, Ann Arbor, MI 48109-1040}
\affiliation{Department of Electrical Engineering and Computer Science, University of Michigan, \\1301 Beal Avenue, Ann Arbor, 48109-2122}

\author{Herbert G. Winful}
 \email{arrays@umich.edu}
\affiliation{Department of Electrical Engineering and Computer Science, University of Michigan, \\1301 Beal Avenue, Ann Arbor, 48109-2122}

\date{\today}

\begin{abstract}
We present a traveling wave model for a semiconductor diode laser based on quantum wells. The gain model is carefully derived from first principles and implemented with as few phenomenological constants as possible. The transverse energies of the quantum well confined electrons are discretized to automatically capture the effects of spectral and spatial hole burning, gain asymmetry, and the linewidth enhancement factor. We apply this model to semiconductor optical amplifiers and single-section phase-locked lasers. We are able to reproduce the experimental results. The calculated frequency modulated comb shows potential to be a compact, chip-scale comb source without additional external components.
\end{abstract}


\maketitle

\section{Introduction}

Optical frequency combs have had a great impact on the fields of ultrafast and nonlinear optics, frequency metrology, and optical spectroscopy in the past few decades \cite{Cundiff2003}. Frequency combs are useful in many applications, including absolute frequency measurement \cite{Udem1999}, multi-heterodyne spectroscopy \cite{Coddington2008}, optical atomic clocks \cite{Diddams2001}, and arbitrary waveform synthesis \cite{Cundiff2010}. Current methods for comb generation include the mode-locking of Ti:Sapphire laser \cite{Sutter1999} and fiber lasers \cite{Fermann2013}, as well as parametric frequency conversion due to the Kerr nonlinearity in passive microresonators \cite{Herr2012}. These approaches, however, require many discrete optical or fiber components, careful alignment, and bulky pump lasers and amplifiers, thus limiting their general utility outside of laboratories. There is thus a need for portable, efficient, robust, and chip-scale comb sources that can be deployed in the field and greatly extend the usefulness of frequency combs. 

Mode locked diode lasers offer the possibility of direct generation of frequency combs from a chip-scale device \cite{Moskalenko2017, Rosales2011}. Typically, passively mode-locked diode lasers comprise two sections: a gain section and a reverse-biased saturable absorber section that leads to the formation of a periodic train of short pulses and hence a comb in the frequency domain. The major obstacle in generating short pulses in diode lasers stems from the nonlinear phase shifts that occur due to fast carrier dynamics \cite{Delfyett1992}, essentially limiting the pulse width inside the cavity. However, single-section diode lasers without saturable absorbers can also operate in a multimode phase-synchronized state known as frequency-modulated (FM) mode locking \cite{Tiemeijer1989}. In the ideal FM mode locked state, the output is a continuous wave in time but the frequency modulation results in a set of comb lines with a fixed, non-zero phase difference. Such FM modelocked operation has been studied most intensively in quantum dot (QD) \cite{Gioannini2015, Rosales2012} and quantum dash \cite{Rosales2012-2} (QDash) lasers, but has also been observed in quantum well (QW) \cite{Sato2003, Calo2015} and bulk semiconductor lasers \cite{Tiemeijer1989}. While some theoretical work has been done for how these combs emerge in a QD single-section laser \cite{Gioannini2015}, a detailed model for FM comb generation in QW diode lasers is still lacking.

There have been many models published for semiconductor quantum well lasers with varying degrees of complexity. The simplest models include only a single rate equation and photon density variable \cite{Homar1996, Arakawa1986}, while more complex models may use multiple rate equations and more complex forms of the material polarization \cite{KN2010, McDonald1995, Jones1995, Vandermeer2005, Gordon2008, Lenstra2014} with varying degrees of phenomenological expressions and constants inserted. However, the existing models are usually insufficiently detailed to explain why FM combs arise in some QW lasers and not others, nor do they indicate which parameters need to be optimized for comb generation. The difficulty in modeling these types of diode lasers stems from the many nonlinear effects in the semiconductor laser cavity that must be properly accounted for. 

In this paper we present a detailed traveling wave model of Fabry-Perot QW diode lasers that elucidates the origin of FM self-mode locking and the formation of frequency combs in these lasers. The model takes into account the multiple cavity modes as a modulation of the electric field envelope, spectral and spatial hole burning, carrier induced refractive index shift, some intraband carrier dynamics, and cavity dispersion. The gain is derived from first principles, starting from the modified Semiconductor Bloch equations with carrier-carrier interactions described through rate equations.  Our approach follows that of previous works \cite{Chow2002, Gioannini2015} but tailored to quantum well nanostructures.  

\section{Theoretical Model}

\begin{figure}[h]
\begin{center}
\includegraphics[scale=1.5]{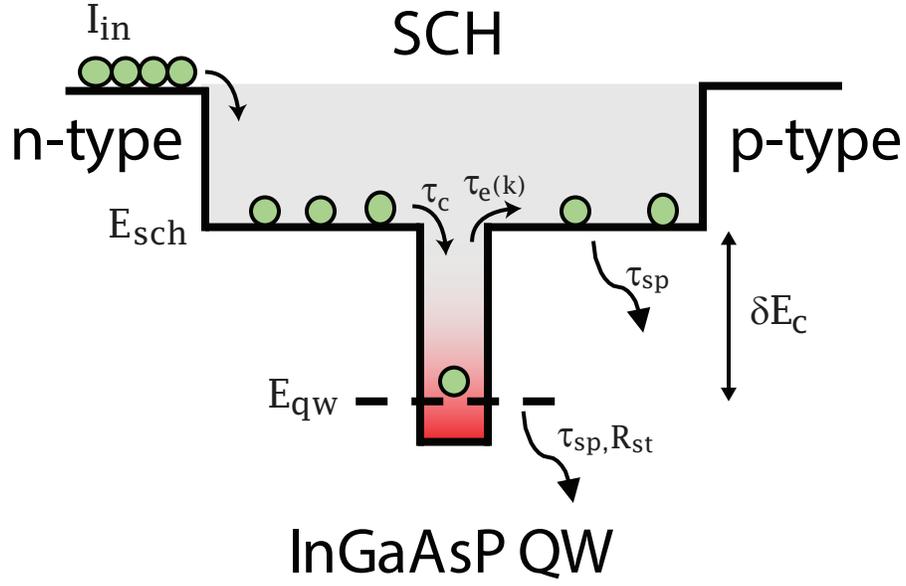}
\end{center}
\caption{A schematic of the quantum well laser diode with current injection into the SCH layer which is captured via $\tau_c$ into the quantum well. The captured electrons have a distribution of transverse energies that can escape the quantum well via $\tau_e(\v{k})$.}
\label{schematic_FIG}
\end{figure}

We start by giving an overview of our model from a physical perspective and write down only the essential equations to be solved while the detailed mathematical derivation is relegated to the appendices. The basic schematic for the model is shown in Figure \ref{schematic_FIG}. Electrons injected from the n side (holes from the p side) drop down to the separate confinement heterostructure (SCH) layer, and become trapped in the quantum well. The most important difference between our quantum well model and previous models is that, for the carriers trapped in the quantum well, we have discretized the carrier equations in energy space and combined them with a truly multimode wave equation. While this approach does increase the number of carrier equations to solve, it captures all the important dynamics of the multiple Fabry-Perot cavity modes and their interactions with carriers at different transverse energies. 

In a semiconductor, the carriers are typically confined in some type of nanostructure, such as a 2-D quantum well, a 1-D quantum wire or a 0-D quantum dot or dash, with an energy distribution determined by the $N$-dimensional density of states $D_r^{N-D}$ and occupation probability for electrons ($e$) or holes ($h$) $\rho^{e,h}$. We assume that the microscopic coherence decays sufficiently quickly such that each individual carrier emits light in a characteristic Lorentzian spectral lineshape with a homogenous linewidth $2\Gamma$ as determined by intraband relaxation effects. However, each group of carriers will emit at a different central frequency. In quantum wells in particular, the carriers have momenta in the unconfined directions that we quantify as the transverse energy $E_t$, and it is these energies that modify the transition frequency for all carriers with energy $E_t$ \cite{Chuang2009}. By integrating all carrier Lorentzians in energy space for each quantum well confined state, we have a gain term that accounts for homogenous and inhomogenous broadening, the asymmetric nature of the gain due to occupation levels and density of states, and the carrier-induced refractive index change. These complex Lorentzians also offer a simple way to calculate the real and imaginary parts of the gain without resorting to the Kramers-Kronig relations.

The electric field of the light wave in the cavity is taken as a sum of forward and backward components

\begin{align}
E(z,t) = E_+(z,t) e^{ik_0 z}+E_-(z,t)e^{-i k_0 z}
\end{align}
whose amplitudes satisfy the slowly-varying envelope equation

\begin{align}
\label{tw_eqn}
\pm\pd{}{z}E_\pm(z,t) + \frac{1}{v_g} \pd{}{t}E_\pm(z,t) = \Gamma_{xy} \frac{\omega_0^2}{2i k_0 c^2 \epsilon_0} \langle P_{tot}(t) e^{\mp ik_0z} \rangle
\end{align}
where the angular brackets signify averaging over a few wavelengths. Here, $v_g = c/n_0$ is the group velocity, $n_0$ is the group refractive index, $\Gamma_{xy}$ is the transverse confinement factor, $\omega_0$ is the central photon frequency (the choice of $\omega_0$ can be arbitrary but is generally chosen to be the transition frequency at the band edge), and $k_0 = n_0 \omega_0/c$. The material polarization $P_{tot}$ is obtained from the Bloch Equations as tailored to semiconductors \cite{ChowKochSargent1994}:

\begin{subequations}
\begin{align}
\label{sc_bloch_p}
i \hbar \pd{p(\v{k},t)}{t} = (\hbar\omega_0 - \Delta E_{cv}(\v{k}) )p(\v{k},t) - \frac{d_{cv} }{2}E(\v{k}, z,t)(\rho^e(\v{k},t) + \rho^h(\v{k},t) - 1)-i\hbar \frac{p(\v{k}, t)}{T_2}
\end{align}
\begin{align}
\label{sc_bloch_ne}
\pd{\rho^e(\v{k},t)}{t} &= -\frac{1}{\hbar} Im[d^*_{cv} E^*(z,t) p(\v{k},t)] + \pd{\rho^e(\v{k},t)}{t}|_{relax}\\
\label{sc_bloch_nh}
\pd{\rho^h(\v{k},t)}{t} &= -\frac{1}{\hbar} Im[d^*_{cv} E^*(z,t) p(\v{k},t)] + \pd{\rho^h(\v{k},t)}{t}|_{relax}
\end{align}
\end{subequations}
where $p(\v{k},t)$ is the microscopic polarization, $\rho^{e,h}(\bf{k},t)$ is the occupation probability of electrons and holes, $d_{cv}$ is the dipole matrix element, $\Delta E_{cv}(\bf{k})$ is the transition energy between the conduction and valence bands, and $T_2 = 1/\Gamma$ is the intraband relaxation time which gives rise to homogenous broadening. It is important to note that these equations are in the time domain but are parameterized by the wavevector $\bf{k}$ and hence represent the time evolution of the subset of carriers with momentum $\bf{k}$.

A key simplification in our model is to assume that the intraband scattering is sufficiently fast to warrant the microscopic polarization adiabatically following the changes in carrier population. For modeling ultra-short pulses, this assumption may no longer hold and a full set of polarization equations will need to be solved dynamically. Integrating Eq. \ref{sc_bloch_p}, we obtain a time domain expression for the microscopic polarization in terms of the occupation probabilities and the electric field:

\begin{align}
p(\v{k}, t) = \frac{i d_{cv} E(\v{k},z,t)}{2\hbar} \int_{-\infty}^t dt' E(z,t') e^{ -\left(i\frac{\Delta E_{cv}(\v{k})}{\hbar}-\omega_0 \right)(t-t')-\Gamma(t-t')} (\rho^e(\v{k},t') + \rho^h(\v{k},t') - 1)
\end{align}
Next, we make the standard adiabatic approximation in which we assume the occupation probabilities evolve slowly compared to the intraband relaxation time $1/\Gamma$ and can be taken out of the integral, with $t'$ replaced by $t$. The remaining convolution integral is then defined as the filtered field \cite{Gioannini2015}

\begin{align}
\label{F_field}
F(\v{k}, z,t) = \Gamma \int_{-\infty}^t dt' e^{i(\frac{\Delta E_{cv}(\v{k}) }{\hbar}-\omega_0)(t-t')-\Gamma(t-t')} E(z,t')
\end{align}

The filtered field consists of all the components that interact with the population $\rho^{e,h}(\bf{k},t)$. Here the transition frequency is defined such that $\hbar\omega_0$ is the transition energy for a confined electron-hole pair with zero transverse energy and satisfies

\begin{align*}
\frac{\Delta E_{cv}(\v{k}) }{\hbar}-\omega_0 = \frac{E_t(\v{k})}{\hbar}
\end{align*}
Thus each discretized carrier group will have a different filtering frequency defined by the transverse energy $E_t$. The time-dependent microscopic polarization reduces to a simple expression:

\begin{align}
\label{p_ss}
p(\v{k}, t) = \frac{i d_{cv}}{2\hbar \Gamma} F(\v{k},z,t)
\end{align}
Here we note that physically, the $\bf{k}$ dependence of the confined carriers in the quantum well is due to a momentum $\bf{k}$ in the two transverse directions, and we therefore define a transverse energy with a simple parabolic band structure:

\begin{align}
E_t &= \frac{\hbar^2 |\v{k}|^2}{2 m_r^*}
\end{align}

\noindent where $m^*_r$ is the reduced effective mass. Hence to save space, we interchangeably write $\rho^{e,h}(\v{k},t) \leftrightarrow \rho^{e,h}_{E_t}$. We can also rewrite the filtered field by interchanging $F(\v{k},z,t) \leftrightarrow F(E_t,z,t)$. 

The total polarization per volume is a summation over all carrier groups with momentum $\bf{k}$. Therefore, the total polarization for a 2-D quantum well can be written:

\begin{align}
\label{ptot_sumk}
P_{tot}(t) = \frac{2}{V} \sum_{\v{k}} d^*_{ev} p(\v{k},t) = i\frac{|d_{cv}|^2}{2\hbar\Gamma} \frac{2}{V} \sum_{\v{k}} (\rho^e_{E_t}+\rho^h_{E_t}-1) F(E_t,z,t)
\end{align}

\noindent The $\v{k}$-summation can be converted to a transverse energy integral. We use a simple parabolic dispersion relation for the conduction and valence bands:

\begin{subequations}
\begin{align}
E_c &= E_g+ E_{e1}+\frac{\hbar^2 |\v{k}|^2}{2 m_e^*}\\
E_v &= E_{h1}-\frac{\hbar^2 |\v{k}|^2}{2 m_h^*}\\
\hbar\omega_0 &= E_g+E_{e1} -E_{h1}
\end{align}
\end{subequations}

\noindent where $E_g$ is the band gap energy, $E_{e1}$ is the confined electron energy, $E_{h1}$ is the confined hole energy, $m^*_{e,h}$ is the electron and hole effective mass (we have assumed only a single confined electron state). Rewriting Eq. \ref{ptot_sumk} with an energy integral, we obtain:

\begin{align}
\label{ptot_t}
P_{tot}(t) =  i\frac{|d_{cv}|^2}{2\hbar\Gamma} \int dE_t D_r^{2D} (\rho^e_{E_t}+\rho^h_{E_t}-1) F(E_t,z,t)
\end{align}

The dipole matrix element can be rewritten as the momentum matrix element via $|d_{cv}|^2 = \frac{q^2}{m_0^2 \omega_0^2} |\uv{e}\cdot \v{p}|^2$ where $q$ is the electron charge and $m_0$ the electron mass. The macroscopic polarization calculated in Eq. \ref{ptot_t} serves as a source term for the forward and backward propagating electric fields in the laser.  The constants on the RHS of \ref{tw_eqn} can be combined to yield a gain coefficient 

\begin{align*}
g_0 &=  \frac{\Gamma_{xy}q^2 D^{2D}_r |\uv{e}_j \cdot \v{p}_{cv}|^2}{2 n_0 c \epsilon_0 m_0^2\Gamma}
\end{align*}

To complete the derivation of the propagation equations, we include the effects of carrier gratings resulting from the interference between forward and backward waves. Our approach to modeling this spatial hole burning (SHB) is to follow the techniques of \cite{Homar1996}, \cite{Javaloyes2009} and \cite{Homar1996-2} and expand the QW population into its second harmonic in space. In this formulation, the population becomes

\begin{align}
\label{p_zexp}
\rho^{e,h}_{E_t} = \rho^{e,h}_{qw, E_t} + \rho_{g,E_t} e^{i2k_0z} + \rho^*_{g,E_t} e^{-i2k_0z} + ...
\end{align}
For simplicity, we have used a single variable for the carrier gratings for both electrons and holes. The filtered  field in the polarization also consists of forward and backward components:

\begin{align}
\label{F_zexp}
F &= F_+ e^{-ik_0 z} + F_- e^{ik_0 z}
\end{align}

Inserting Eqs. \ref{ptot_t} \ref{p_zexp} \ref{F_zexp} in Eq. \ref{tw_eqn} and keeping only the phase-matched terms we obtain the electric field equations:

\begin{align}
\label{wave_eq_simple}
\begin{split}
\pm\pd{E_\pm}{z}+ &\frac{1}{v_g} \pd{E_\pm}{t} = \frac{g_0}{2} \int \frac{dE_t}{\hbar\omega_0} (\rho^e_{qw, E_t}+\rho^h_{qw,E_t}-1) F_\pm(E_t, z,t)\\
 & + g_0 \int \frac{dE_t}{\hbar\omega_0} \rho^{(*)}_{g,E_t} F_\mp(E_t, z,t)
\end{split}
\end{align}

We note that the grating term $\rho_{g,E_t}^{(*)}$ is associated with the forward wave equation and its conjugate with the backward wave. Finally, we simply add the additional terms in Eq. \ref{wave_eq_simple} that describe standard linear and nonlinear effects, and scale via $n_{qw}$, the number of quantum wells to obtain                                                                                                                                 

\begin{align}
\begin{split}
\label{wave_eq}
\pm \pd{E_\pm}{z}+ &\frac{1}{v_g} \pd{E_\pm}{t} + i\frac{k''}{2} \pdd{E_\pm}{t} =  -\frac{\alpha}{2} E_\pm - \left(\frac{\alpha_S}{2}+i\beta_S\right)(|E_\pm|^2+2|E_\mp|^2)E_\pm +S_{sp} \\
 & + n_{qw} \frac{g_0}{2} \int \frac{dE_t}{\hbar\omega_0} (\rho^e_{qw, E_t}+\rho^h_{qw,E_t}-1) F_\pm(E_t, z,t)\\
 & + n_{qw} g_0 \int \frac{dE_t}{\hbar\omega_0} \rho^{(*)}_{g,E_t} F_\mp(E_t, z,t)
\end{split}
\end{align}

\noindent where $k''$ is the dispersion coefficient, $\alpha$ is the linear waveguide loss, and $\alpha_S, \beta_S$ are respectively the two-photon absorption and Kerr nonlinear coefficients, and $S_{sp}$ is the spontaneous emission term derived in the Appendix.

These field equations are coupled with the carrier rate equations for the SCH and QW sections. The QW equations are labeled with the transverse variable  for each discretized bin yielding

\begin{align}
\label{sch_eq}
\pd{\rho^{e,h}_{sch}}{t} = \frac{\eta J_{in}}{q N_{c,v,sch}h_{sch}}(1-\rho^{e,h}_{sch})- \frac{\rho^{e,h}_{sch}}{\tau_{sp}} +n_{qw}\sum_{E_t} \left[\rho^{e,h}_{qw, E_t}\frac{(1-\rho^{e,h}_{sch})}{\tau^{e,h}_e}-\rho^{e,h}_{sch}\frac{(1-\rho^{e,h}_{qw,E_t})}{\tau^{e,h}_c}\right]\\
\label{qw_eq}
\pd{\rho^{e,h}_{qw, E_t}}{t} = \frac{h_{sch}N_{c,v,sch}}{n_{qw} h_{qw}N_{r,qw}}\left(\rho^{e,h}_{sch}\frac{(1-\rho^{e,h}_{qw,E_t})}{\tau^{e,h}_c} - \rho^{e,h}_{qw,E_t}\frac{(1-\rho^{e,h}_{sch})}{\tau^{e,h}_e}\right) - \frac{\rho^{e,h}_{qw, E_t}}{\tau_{sp}} - R_{st} - R_{g}
\end{align}

\begin{align}
\begin{split}
\label{pg_eq}
\pd{\rho_{g, E_t}}{t} &=- \frac{\rho_{g,E_t}}{\tau_{sp}} -4k_0^2 D \rho_{g,E_t} - 2g_0 \frac{\Delta E_t}{(\hbar\omega_0)^2 h_{qw} W N_{r,qw}}\\
&\times \left[ \frac{1}{2}(E_+^*F_- + F_+^* E_-)  (\rho^e_{qw}+\rho^h_{qw}-1) + 2\text{Re}(E^*_+F_+ + E^*_-F_-)\rho_{g,E_t}\right]
\end{split}
\end{align}

\begin{align}
R_{st} &= 2g_0 \frac{\Delta E_t}{(\hbar\omega_0)^2 h_{qw} W N_{r,qw}} (\rho_{qw, E_t}^e+\rho^h_{qw, E_t}-1) \text{Re}(E^* F)\\
R_g &= 2g_0 \frac{\Delta E_t}{(\hbar\omega_0)^2 h_{qw} W N_{r,qw}} \big((E_+F^*_- + F_+ E^*_-)\rho_{g,E_t} + (E_+^*F_- + F_+^* E_-)\rho^*_{g,E_t}\big)
\end{align}

\noindent where $N_{c,v,sch} = 2 \left(\frac{m_{e,h}^* k_B T}{2\hbar^2 \pi}\right)^{3/2}$, $N_r = \frac{m_{r}^* \Delta E_t }{\hbar^2 \pi h_{qw}}$ are the effective 3-D and 2-D density of states, $D$ is the ambipolar diffusion coefficient, $\tau_{sp}$ is the spontaneous emission lifetime, $\tau^{e,h}_c$ is the capture lifetime, and $\tau^{e,h}_e$ is the escape lifetime. The recombination rates $R_{st}$ and $R_g$ govern population decay due to stimulated emission and the carrier grating respectively. The escape times $\tau_e^{e,h}$ are particularly important in our model as they phenomenologically represent intraband interactions.  As shown in the Appendix, they are given by 

\begin{align}
\tau^e_e = \tau^e_c \exp( (\delta E_c-\frac{m^*_r}{m^*_e}E_t)/k_B T)\\
\tau^h_e = \tau^h_c \exp( (\delta E_v-\frac{m^*_r}{m^*_h}E_t)/k_B T)
\end{align}
The value of these escape times is tailored specifically to allow the rate equations \ref{sch_eq}, \ref{qw_eq} to relax to the Fermi-Dirac distribution.

\section{Numerical Results for Pulse Amplification}

We solve the forward and backward wave equations (Eq. \ref{wave_eq}), coupled with the carrier rate equations (Eqs. \ref{sch_eq},\ref{qw_eq},\ref{pg_eq}) numerically using a first order Euler scheme very similar to reference \cite{Rossetti2011}. We have chosen a time step of $\Delta t = 30$ fs with the full simulation parameters listed in Table \ref{material_param}. 

In order to solve the full set of equations, one must specify the limits to $E_t$ as well as the number of different energy bins. The maximum transverse energy can be set to the quantum well barrier height, as there will not be any confined carriers with a total energy that surpasses this value. However, the maximum can be lower if the pump current is not too large, as then the high energy carriers will not significantly contribute to the total gain. For our simulations, we have chosen the values max$(E_t) = 50$ meV with 25 energy bins for a total of 75 quantum well carrier equations (25 for both electron and hole equations, 25 for the grating term). The energy step $\Delta E_t = 2$ meV is small relative to the homogenous FWHM (2$\Gamma$) to ensure reasonable accuracy in the gain integral. 

\begin{figure}[h]
\begin{center}
\includegraphics[scale=1.0]{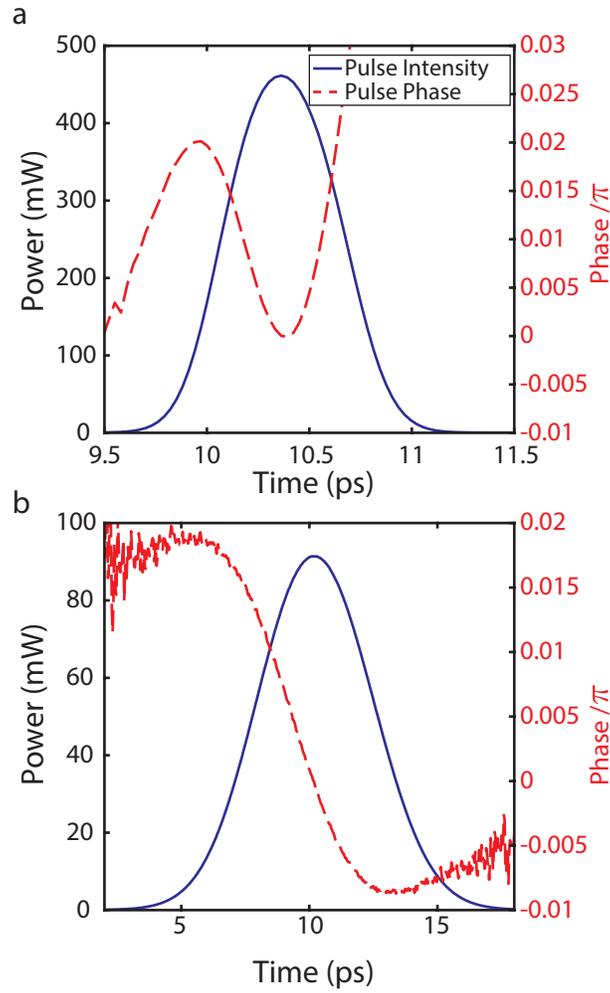}
\end{center}
\caption{The results of sending an unchirped gaussian pulse through a single pass of the laser cavity. The amplified pulse shape and phase are plotted for a) a 0.5 ps pulse b) a 5 ps pulse.}
\label{pulses_FIG}
\end{figure}

\begin{figure}[h]
\begin{center}
\includegraphics[scale=1.0]{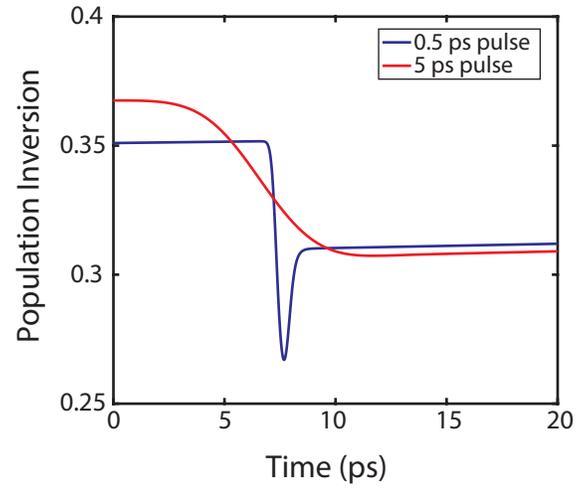}
\end{center}
\caption{The population depletion and recovery as the pulses pass through. }
\label{pinv_FIG}
\end{figure}

We first solve the equations for a pulse passing once through the laser cavity without facet reflections, acting as a semiconductor optical amplifier (SOA) in order to test that the phase shifts are accurately modeled. It is important to model these phase distortions accurately, as they typically work against mode locking. The results of our simulation are shown in Figure \ref{pulses_FIG}. The pulse phase varies as expected, which for the long pulse (5 ps) resembles a more linear shape while the shorter pulse (0.5 ps) retains a cubic shape due to the carrier induced refractive index change. The population depletion and recovery, shown in Figure \ref{pinv_FIG} are consistent in behavior with results from simpler impulse response models \cite{Delfyett1992}. For the long pulse (5 ps), the population depletion is mostly monotonic and follows a smooth curve. However, a short pulse (0.5 ps) will deplete the population quickly but the gain will partially recover due to carrier cooling. These fast carrier dynamics are the primary cause of the cubic phase shifts in the amplified pulse and are detrimental to the generation of ultra short pulses. In our simulations, carrier cooling occurs as additional carriers drop down from the SCH layer to fill the vacant QW states depleted by the short pulse. This capture time is on the order of a picosecond, thus only pulses much shorter than this time will see the effects of carrier cooling. These results verify the accuracy of our gain calculations with previous pulse amplification experiments \cite{Delfyett1992}. 

\section{Numerical Results for a Diode Laser}

\begin{figure}[h]
\begin{center}
\includegraphics[scale=1]{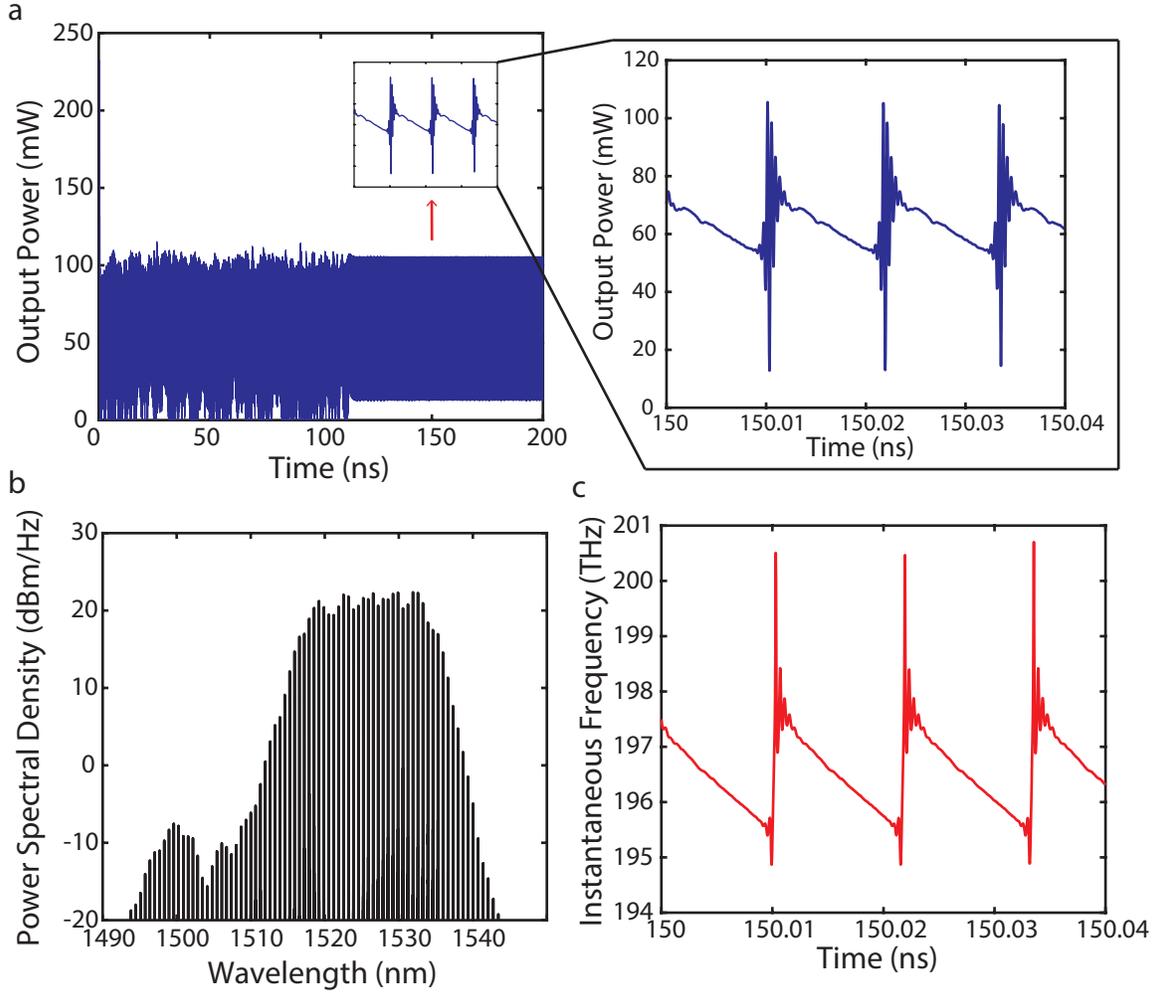}
\end{center}
\caption{a) The temporal output of the single-section quantum well device at $\eta I_{in} = 25$mA with a zoomed inset to show the detailed dynamics. The output is quasi-CW except for a short burst that repeats every round trip. A steady state is reached for $t > 110$ ns b) The power spectral density of the last 100 ns of the temporal output in log scale showing a broad comb c) the instantaneous frequency of the laser output, which is also sweeping periodically, showing the FM nature of the comb.}
\label{combI_FIG}
\end{figure}

\begin{figure}[h]
\begin{center}
\includegraphics[scale=1]{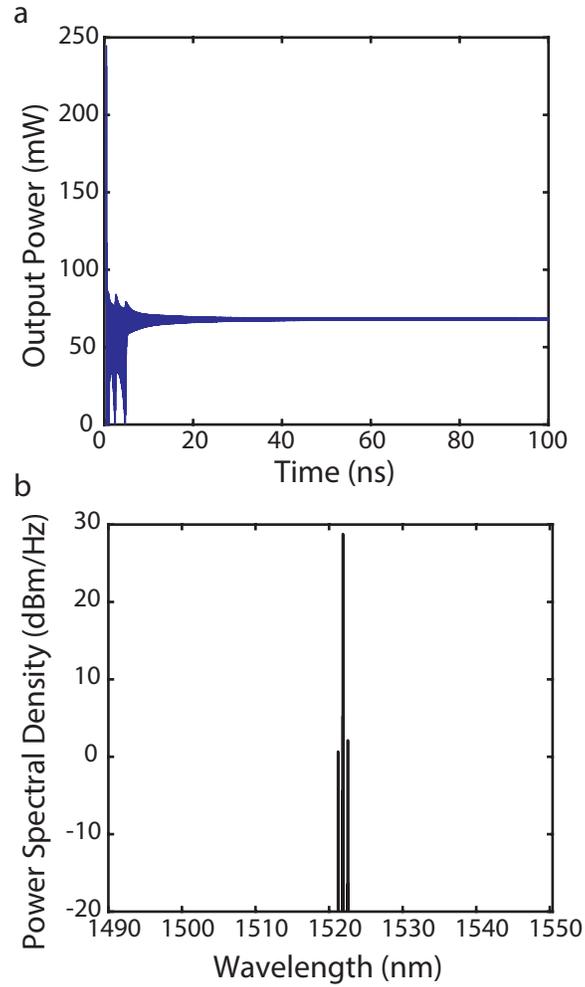}
\end{center}
\caption{a) the temporal output of the laser with the population grating term set to zero. The output relaxes to a single mode after some time b) the spectrum of the above output which shows a single mode dominating, in stark contrast to the case when the grating is on (Figure \ref{combI_FIG}b).}
\label{nog_FIG}
\end{figure}

While we have successfully modeled the phase dynamics in single-pass pulse amplification, the primary application of our model is to investigate FM frequency comb generation in a single-section diode laser. We simulate 200 ns of a cleaved facet laser starting from noise and monitor the output. The results are plotted in Figure \ref{combI_FIG}. The temporal output and spectrum match well with the experimental results for a single-section laser found in references \cite{Sato2003} and \cite{Calo2015}, with a significant number of strong comb lines spanning about 30 nm in bandwidth with a mode spacing of $\nu_{fsr} = 85.7$ GHz. There is much irregular oscillation in the temporal output until steady state is reached. After steady state ($t  > 110$ns), the waveform remains periodic and is coherent over a long timespan. The output also does not consist of a train of short pulses, but rather a periodic modulation of the intensity and phase to generate the comb spectrum, as seen in the zoomed in plots of the output power and instantaneous frequency in Figure \ref{combI_FIG}a,c. The frequency is periodic, being swept across a large range of about 5 THz. The primary mechanism for the generation of multiple Fabry-Perot modes is the spatial hole burning grating term $\rho_{g,E_t}$, consistent with previous work \cite{Gioannini2015}, \cite{Javaloyes2009}. This term allows several modes to lase at once and acts as a conduit for four-wave mixing. We verify this by turning off the grating term and we only obtain a single lasing mode after the initial relaxations as shown in Figure \ref{nog_FIG}. 

\begin{figure}[h]
\begin{center}
\includegraphics[scale=1]{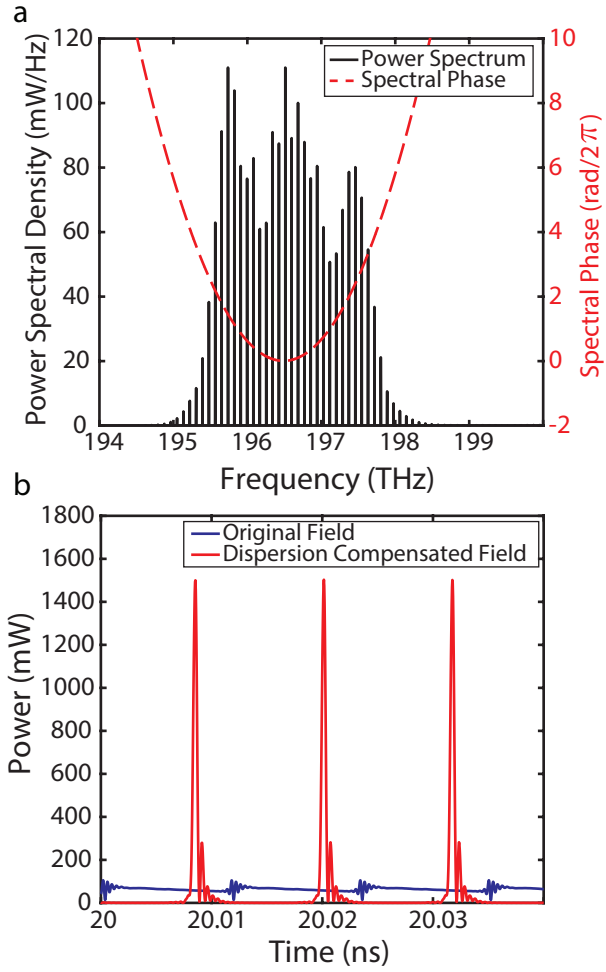}
\end{center}
\caption{a) the power spectral density in linear scale along with the spectral phase b) The spectrum is compensated for dispersion and inverse Fourier transformed to produce a series of short pulses separated by the cavity round trip time. The group delay dispersion is calculated to be 0.41 ps$^2$.}
\label{dc_FIG}
\end{figure}

To show that the modes are indeed locked, we plot the spectrum and spectral phase in linear scale in Figure \ref{dc_FIG}a. This quadratic phase can be compensated by propagation through anomalous dispersion fibers \cite{Sato2003}, transforming the output into a series of short pulses. We simulate this compensation by multiplying our spectrum by the transfer function $H(\omega) = e^{- i \text{GDD} \omega^2}$\cite{Gioannini2015}  where GDD is the group delay dispersion, calculated to be $0.41$ ps$^2$. After applying the inverse Fourier transform, we see a series of short pulses ($\approx$ 390 fs FWHM) emerge, which is indicative of mode locking\cite{Rosales2012-2}. The original field and dispersion compensated field are plotted in Figure \ref{dc_FIG}b for comparison. We note that the compensation is not perfect, as there is a small side pulse in front of the main pulse that indicates that the output pulses have higher order chirp that is not compensated by the simple application of quadratic phase \cite{Gioannini2015}. However, the fact that the phase compensation can result in a series of short pulses suggests the field inside the cavity is actually a train of highly chirped pulses.

In order for this comb to be practical, the linewidth of each mode must be very narrow for many of the high resolution comb spectroscopy techniques to be used. Unfortunately we could not obtain an exact value for the linewidth of our comb as an accurate measurement requires a very lengthy sample of data in the time domain, which is difficult to obtain from a computational standpoint. We have run simulations up to 1.5 $\mu$s and attempted to measure the linewidth but even at such time scales, the linewidth was still limited by the time window. Despite this, we calculate an upper bound of 1 MHz for the linewidth, while the real RF linewidth may be much smaller in the tens of kHz range \cite{Calo2015}. 

\section{Discussion}

The results shown in Figures \ref{combI_FIG}, \ref{dc_FIG} show that single-section QW diode lasers have the potential to produce useful frequency combs. The FM nature of the comb and the ability to convert FM into a series of pulses via external dispersion compensation may prove useful for probing either fragile samples that require low intensity or samples that benefit from high pulse power. Moreover, the planar processes used in manufacturing such diodes are well developed and allow many lasers to be made at once. While the bandwidth is already sufficiently large, a wider bandwidth may be achieved by combining several lasers together, each with an offset to the central lasing frequency by adjusting the bandgap of the semiconductor material. The mode spacing can also be adjusted by changing the length of these lasers anywhere from a few hundred microns to several millimeters, and perhaps even on a finer scale by adjusting the pump current \cite{Sato2003} for multiheterodyne measurements. Because the entire comb is generated on the chip itself without any external mirrors or components, the single-section QW diode laser represents a highly portable source of frequency combs.

We have found that several material parameters are vital to the generation of FM combs. First, the homogenous linewidth $2\Gamma$ should be reasonably large compared to the mode spacing, primarily to facilitate strong four-wave mixing (FWM) interactions to lock the modes together. In addition, too small of a homogenous linewidth may allow additional modes to lase independently from decreased gain competition, with additional gain coming from the inhomogenously distributed carriers. Once this occurs, there is no mechanism for locking these disparate modes, as FWM is no longer effective due to these modes falling outside of the homogenous linewidth. Second, for effective multimode lasing, we need a SHB effect, or a low enough diffusion constant, in order to see comb generation. The InGaAsP QW system is well suited to satisfy this requirement, as the laser operates in the near-IR so that the half-wavelength grating spacing exceeds the diffusion length.  Compared to other materials such as GaAs, the quaternary alloy InGaAsP has low measured values of diffusion \cite{Marshall2000}. It is the persistence of the spatially burnt holes that leads to gain suppression \cite{Su1988} as well as multimode operation. 

Moreover, we have found, perhaps surprisingly, that other effects have very little impact on the generation of combs. Second order material and waveguide dispersion, as modeled by the parameter $k''$ , has only a very minor effect on mode locking, as the laser produces an FM mode locked state regardless of the inclusion of dispersion. The third-order Kerr nonlinearity and two- photon absorption also do not significantly alter the FM output, consistent with previous findings in QD systems \cite{Gioannini2015}.

We have used typical values for many of the physical parameters appropriate to an InGaAsP system and we see these combs emerge naturally through spatial hole burning and four-wave mixing. However, the interaction of the various physical phenomena is rather complex and we will present a more thorough study on the physics behind these combs in a future work.

\section{Conclusion and Acknowledgments}

In conclusion, we have presented a comprehensive traveling wave model for a quantum-well based semiconductor laser. We have validated the accuracy of the calculations by replicating a few experimental results, particularly generating frequency combs from single-section diode lasers. This model should serve as a suitable platform for additional studies into the physics that enables these combs to be generated and possibly discover new ways to achieve stable mode-locking in these diode lasers. Long-wavelength QW lasers show much promise as practical, chip-scale sources of FM combs with the necessary bandwidth and linewidth for the many applications of frequency combs.

This research was developed with funding from the Defense Advanced Research Projects Agency (DARPA) through the SCOUT program. The views, opinions and/or findings expressed are those of the author and should not be interpreted as representing the official views or policies of the Department of Defense or the U.S. Government. This research was also supported in part through computational resources and services provided by Advanced Research Computing at the University of Michigan, Ann Arbor.

\appendix
\section{Derivation of Rate Equations}

Use of the microscopic polarization Eq.  \ref{p_ss}  in Eqs. \ref{sc_bloch_ne}, \ref{sc_bloch_nh} yields the rate equations

\begin{align}
\pd{\rho^{e,h}(\v{k},t)}{t} &= -\frac{|d_{cv}|^2}{2\hbar^2\Gamma} (\rho^e(\v{k},t)+\rho^h(\v{k},t)-1) \text{Re}(E^* F) + \pd{\rho^{e,h}(\v{k},t)}{t}|_{relax}
\end{align}

We rewrite the electric fields in a convenient form: the electric field is scaled to be in units of $\sqrt{\text{Watts}}$ via the expression $|E(z,t)|^2 \rightarrow \frac{2 \Gamma_{xy}}{h_{qw} W c n_0 \epsilon_0} |E(z,t)|^2$, where $\Gamma_{xy}$ is the confinement factor, $h_{qw}$ is the height of the quantum well and $W$ is the width. 

\begin{align}
\pd{\rho^{e,h}(\v{k},t)}{t} &= -\frac{\Gamma_{xy} |\uv{e}_j \cdot \v{p}_{cv}|^2 q^2 }{n_0 c \epsilon_0 m_0^2 \Gamma} \frac{1}{(\hbar\omega_0)^2 h_{qw} W} (\rho^e(\v{k},t)+\rho^h(\v{k},t)-1) \text{Re}(E^* F) + \pd{\rho^{e,h}(\v{k},t)}{t}|_{relax}
\end{align}

\noindent We define the gain coefficient $g_0$ and the energy-discretized, reduced density of states, $N_{r,qw}=\Delta E_t D^{2D}_r$ and rewrite the rate equations:

\begin{align}
\label{rho_eq_s}
\pd{\rho^{e,h}(\v{k},t)}{t} &= -2g_0 \frac{\Delta E_t}{(\hbar\omega_0)^2 h_{qw} W N_{r,qw}} (\rho^e(\v{k},t)+\rho^h(\v{k},t)-1) \text{Re}(E^* F) + \pd{\rho^{e,h}(\v{k},t)}{t}|_{relax}
\end{align}

So far, we have only applied a two-level approach to the rate equations even though a semiconductor is actually a four-level system \cite{ChowKochSargent1994}. However, because we solve the electron and holes separately based on the input current and charge conservation, we allow for the cases in which an electron may exist but no hole, and vice versa. In this case, the occupation probabilities $\rho^{e,h}$ no longer obey two-level relations ($\rho^{e}-\rho^{h} = 0$) but can take on any value between 0 and 1 according on the relaxation and pump terms. 

\section{Evaluation of Carrier Relaxation Terms}

In order to progress further, we need to evaluate the electron and hole relaxation terms $\pd{\rho^{e,h}(\v{k},t)}{t}|_{relax}$. We follow the capture and escape approach presented in \cite{Gioannini2015}. First, we start with simple rate equations (without the presence of photons) for carrier number in the SCH and QW layers that satisfy charge conservation.

\begin{subequations}
\begin{align}
\d{N_{sch}}{t} &= -\frac{N_{sch}}{\tau_c}+\frac{N_{qw}}{\tau_e}\\
\d{N_{qw}}{t} &= \frac{N_{sch}}{\tau_c}-\frac{N_{qw}}{\tau_e}\\
\d{}{t} (N_{sch}&+N_{qw}) = 0
\end{align}
\end{subequations} 
We can convert this to occupation probability equations using the relations

\begin{subequations}
\begin{align}
N_{sch} &= N_{c,v,sch} W h_{sch} \Delta z \rho^{e,h}_{sch}\\
N_{qw} &= N_{r,qw} W h_{qw} \Delta z \rho^{e,h}_{qw,E_t}
\end{align}
\end{subequations} 
We also add Pauli blocking terms, which results in the following differential equations for the occupation probabilities:

\begin{subequations}
\label{rate_relax_eqns}
\begin{align}
\pd{\rho^{e,h}_{sch}}{t} &= -\frac{\rho^{e,h}_{sch}}{\tau^{e,h}_c}(1-\rho^{e,h}_{qw,E_t})+\frac{N_{r,qw} h_{qw}}{N_{c,v,sch} h_{sch}} \frac{\rho^{e,h}_{qw,E_t}}{\tau^{e,h}_e}(1-\rho^{e,h}_{sch})\\
\pd{\rho^{e,h}_{qw,E_t}}{t} &= \frac{N_{c,v,sch} h_{sch}}{N_{r, qw} h_{qw}} \frac{\rho^{e,h}_{sch}}{\tau^{e,h}_c}(1-\rho^{e,h}_{qw,E_t})-\frac{\rho^{e,h}_{qw,E_t}}{\tau^{e,h}_e}(1-\rho^{e,h}_{sch})
\end{align}
\end{subequations}
The steady state solutions to Eqns \ref{rate_relax_eqns} should relax into a Fermi-Dirac distribution. We assume the solutions for the electrons (holes follow a similar expression) are of the form:

\begin{subequations}
\begin{align}
\rho^e_{sch} &= \frac{1}{1+\exp\left( \frac{E_{sch} - E_f}{k_B T} \right)}\\
\rho^e_{qw,E_t} &= \frac{1}{1+\exp\left( \frac{E_{qw}+\frac{m^*_r}{m^*_e} E_t - E_f} {k_B T} \right)}
\end{align}
\end{subequations}
where $E_f$ is the electron Fermi level. We can use these solutions in Eqs. \ref{rate_relax_eqns} and solve for the proper escape time in terms of the capture time such that the occupation probabilities settle into a Fermi-Dirac distribution. We find the resulting expressions for the escape times and the relaxation to be:

\begin{subequations}
\begin{align}
\tau^e_e = \tau^e_c \left (\frac{N_{r,qw} h_{qw}}{N_{c,sch} h_{sch}} \right) \exp( (\delta E_c-\frac{m^*_r}{m^*_e} E_t)/k_B T)\\
\tau^h_e = \tau^h_c \left( \frac{N_{r,qw} h_{qw}}{N_{v,sch} h_{sch}}\right) \exp( (\delta E_v-\frac{m^*_r}{m^*_h} E_t)/k_B T)
\end{align}
\end{subequations}

\noindent Here, $\delta E_c = E_{sch} - E_{qw}$ (and analogously, $\delta E_v$) is the energy difference between the SCH layer and the confined carrier with zero transverse energy, visually labeled in Figure \ref{schematic_FIG}. Lastly, we remove the bracketed fraction and write it explicitly in the rate equations, allowing us to define the escape lifetimes more simply as:

\begin{subequations}
\begin{align}
\tau^e_e = \tau^e_c \exp( (\delta E_c-\frac{m^*_r}{m^*_e} E_t)/k_B T)\\
\tau^h_e = \tau^h_c \exp( (\delta E_v-\frac{m^*_r}{m^*_h} E_t)/k_B T)
\end{align}
\end{subequations}
While we have shown the derivation for only a single quantum well carrier group, there are actually multiple quantum well rate equations. Thus the SCH equation must sum up the capture and escape contributions from every group of quantum well carriers. 

\section{Carrier Grating Terms}

The stimulated emission term in the rate equations contains the product:
\begin{align*}
\text{Re}(E^*F) &= \text{Re}(E^*_+F_+ + E^*_- F_-) + \frac{1}{2}(E_+^*F_- + F_+^* E_-) e^{i2k_0z} + \frac{1}{2}(E_+F^*_- + F_+ E^*_-)e^{-i2k_0z}
\end{align*}
Equating the phase-matched portions of the LHS population expansion and the RHS stimulated emission terms, we obtain two separate equations, one for the CW population and a second for the spatial grating terms. For the grating equation, we have added the diffusion term of the form $-D \pdd{}{z}$ on the RHS, where $D$ is the diffusion coefficient. The resulting equations are

\begin{align}
\begin{split}
\pd{\rho^{e,h}_{qw, E_t}}{t} &= ... - 2g_0 \frac{\Delta E_t}{(\hbar\omega_0)^2 h_{qw} W N_{r,qw}} \times \Big[\text{Re}(E^*_+F_+ + E^*_-F_-)(\rho^e_{qw,E_t}+\rho^h_{qw,E_t}-1)\\
&+ (E_+F^*_- + F_+ E^*_-)\rho_{g,E_t} + (E_+^*F_- + F_+^* E_-)\rho^*_{g,E_t} \Big]
\end{split}
\end{align}

\begin{align}
\begin{split}
\pd{\rho_{g,E_t}}{t} &= - 4k_0^2 D \rho_{g,E_t} - 2g_0 \frac{\Delta E_t}{(\hbar\omega_0)^2 h_{qw} W N_{r,qw}}\\
&\times \left[ \frac{1}{2}(E_+^*F_- + F_+^* E_-)  (\rho^e_{qw,E_t}+\rho^h_{qw,E_t}-1) + 2\text{Re}(E^*_+F_+ + E^*_-F_-)\rho_{g,E_t}\right]
\end{split}
\end{align}
The stimulated emission rate and the photon-grating interaction as are now clearly identified as:

\begin{align*}
R_{st} &= 2g_0 \frac{\Delta E_t}{(\hbar\omega_0)^2 h_{qw} W N_{r,qw}} (\rho^e_{qw,E_t}+\rho^h_{qw,E_t}-1) \text{Re}(E^* F)\\
R_g &= 2g_0 \frac{\Delta E_t}{(\hbar\omega_0)^2 h_{qw} W N_{r,qw}} \big((E_+F^*_- + F_+ E^*_-)\rho_{g,E_t} + (E_+^*F_- + F_+^* E_-)\rho^*_{g,E_t}\big)
\end{align*}
Combining all the elements together and adding in the pump $J_{in} = I_{in}/WL$ and spontaneous emission terms, we have the final form of the rate equations:

\begin{align}
\pd{\rho^{e,h}_{sch}}{t} = \frac{\eta J_{in}}{q N_{c,v,sch}h_{sch}}(1-\rho^{e,h}_{sch})- \frac{\rho^{e,h}_{sch}}{\tau_{sp}} +\sum_{E_t} \left[\rho^{e,h}_{qw, E_t}\frac{(1-\rho^{e,h}_{sch})}{\tau^{e,h}_e}-\rho^{e,h}_{sch}\frac{(1-\rho^{e,h}_{qw,E_t})}{\tau^{e,h}_c}\right]\\
\pd{\rho^{e,h}_{qw, E_t}}{t} = \frac{h_{sch}N_{c,v,sch}}{h_{qw}N_{r,qw}}\left(\rho^{e,h}_{sch}\frac{(1-\rho^{e,h}_{qw,E_t})}{\tau^{e,h}_c} - \rho^{e,h}_{qw,E_t}\frac{(1-\rho^{e,h}_{sch})}{\tau^{e,h}_e}\right) - \frac{\rho^{e,h}_{qw, E_t}}{\tau_{sp}} - R_{st} - R_{g}
\end{align}

\begin{align}
\begin{split}
\pd{\rho_{g, E_t}}{t} &=- \frac{\rho_{g,E_t}}{\tau_{sp}} -4k_0^2 D \rho_{g,E_t} - 2g_0 \frac{\Delta E_t}{(\hbar\omega_0)^2 h_{qw} W N_{r,qw}}\\
&\times \left[ \frac{1}{2}(E_+^*F_- + F_+^* E_-)  (\rho^e_{qw}+\rho^h_{qw}-1) + 2\text{Re}(E^*_+F_+ + E^*_-F_-)\rho_{g,E_t}\right]
\end{split}
\end{align}

\section{Derivation of the Gain Spectrum}

\begin{figure}[h]
\begin{center}
\includegraphics[scale=1]{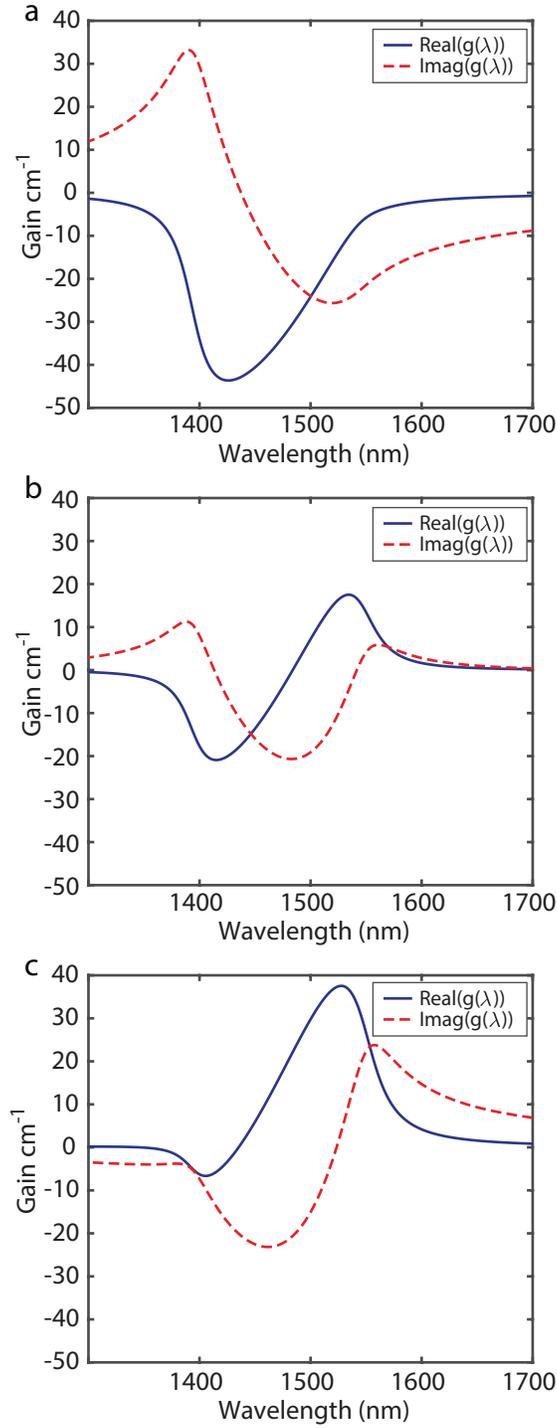}
\end{center}
\caption{The real and imaginary parts of the gain for various levels of carrier injection. a) low injection b) medium injection c) high injection. The gain is asymmetric, reflecting the product of the 2-D density of states and the Fermi-Dirac occupation probabilities.}
\label{gain_FIG}
\end{figure}

We can take a Fourier transform of the gain term in the traveling wave equation in order to visualize the gain spectrum. We assume the carriers are in steady state so that the populations obey Fermi-Dirac statistics. In this case, the Fourier transform evaluates to

\begin{align}
\begin{split}
\mathcal{F} &\left[ \frac{g_0}{2} \int \frac{dE_t}{\hbar\omega_0} (\rho^e_{qw, E_t}+\rho^h_{qw,E_t}-1) F_\pm(E_t, z,t) \right] \\
 &=\frac{g_0}{2} \int \frac{dE_t}{\hbar\omega_0} (\rho^e_{qw, E_t}+\rho^h_{qw,E_t}-1) \frac{E_\pm(z,\omega)}{-i(\omega+E_t/\hbar)-\Gamma}
\end{split}
\end{align}
and hence the field gain is

\begin{align}
\label{g_w}
g(\omega) =\frac{g_0}{2} \int \frac{dE_t}{\hbar\omega_0} (\rho^e_{qw, E_t}+\rho^h_{qw,E_t}-1) \frac{1}{-i(\omega+E_t/\hbar)-\Gamma}
\end{align}
In this form, we see that the gain spectrum consists of a series of Lorentzians centered at different transition energies. A plot of Eq. \ref{g_w} is shown in Figure \ref{gain_FIG} for varying levels of carrier population. 

\section{Derivation of Spontaneous Emission}

Lastly, the spontaneous emission term $S_{sp}$ is derived more phenomenologically. The spontaneous emission term is found by following the approach in \cite{Rossetti2011} in which the power spectrum follows the quantum well gain spectrum. 

\begin{align*}
|S_{sp}\Delta z|^2 &= \sum_{modes} \frac{\text{\# carriers}}{\tau_{sp}} \times \text{photon energy} \times \text{coupling factor}\\
&=  \sum_{modes} n_{qw} \frac{N_{r,qw} h_{qw} W \Delta z}{2\pi \tau_{sp}} \rho^e_{qw,E_t}\rho^h_{qw,E_t} \hbar\omega \beta_{sp}
\end{align*}

\begin{align}
S_{sp} \approx  \sum_{E_t} \sqrt{ \frac{n_{qw} N_{r,qw} h_{qw} W \beta_{sp} \hbar\omega \rho^e_{qw,E_t}\rho^h_{qw,E_t} }{2\pi \tau_{sp} \Delta z}} F_{sp}(E_t) \\
F_{sp}(E_t) = \Gamma \int_{-\infty}^t dt' e^{i(\frac{\Delta E_{cv}}{\hbar}-\omega_0)(t-t')-\Gamma(t-t')} e^{i \phi(z,t',E_t)}
\end{align}

\noindent Here, $\phi(z,t,E_t)$ is a random phase value between 0 and $2\pi$, and $\Delta z = c \Delta t /n_0 $ is the spatial discretization size.

\bibliography{QWLaser_master}

\providecommand{\noopsort}[1]{}\providecommand{\singleletter}[1]{#1}%
\begin{thebibliography}{33}%
\makeatletter
\providecommand \@ifxundefined [1]{%
 \@ifx{#1\undefined}
}%
\providecommand \@ifnum [1]{%
 \ifnum #1\expandafter \@firstoftwo
 \else \expandafter \@secondoftwo
 \fi
}%
\providecommand \@ifx [1]{%
 \ifx #1\expandafter \@firstoftwo
 \else \expandafter \@secondoftwo
 \fi
}%
\providecommand \natexlab [1]{#1}%
\providecommand \enquote  [1]{``#1''}%
\providecommand \bibnamefont  [1]{#1}%
\providecommand \bibfnamefont [1]{#1}%
\providecommand \citenamefont [1]{#1}%
\providecommand \href@noop [0]{\@secondoftwo}%
\providecommand \href [0]{\begingroup \@sanitize@url \@href}%
\providecommand \@href[1]{\@@startlink{#1}\@@href}%
\providecommand \@@href[1]{\endgroup#1\@@endlink}%
\providecommand \@sanitize@url [0]{\catcode `\\12\catcode `\$12\catcode
  `\&12\catcode `\#12\catcode `\^12\catcode `\_12\catcode `\%12\relax}%
\providecommand \@@startlink[1]{}%
\providecommand \@@endlink[0]{}%
\providecommand \url  [0]{\begingroup\@sanitize@url \@url }%
\providecommand \@url [1]{\endgroup\@href {#1}{\urlprefix }}%
\providecommand \urlprefix  [0]{URL }%
\providecommand \Eprint [0]{\href }%
\providecommand \doibase [0]{http://dx.doi.org/}%
\providecommand \selectlanguage [0]{\@gobble}%
\providecommand \bibinfo  [0]{\@secondoftwo}%
\providecommand \bibfield  [0]{\@secondoftwo}%
\providecommand \translation [1]{[#1]}%
\providecommand \BibitemOpen [0]{}%
\providecommand \bibitemStop [0]{}%
\providecommand \bibitemNoStop [0]{.\EOS\space}%
\providecommand \EOS [0]{\spacefactor3000\relax}%
\providecommand \BibitemShut  [1]{\csname bibitem#1\endcsname}%
\let\auto@bib@innerbib\@empty
%</preamble>
\bibitem [{\citenamefont {Cundiff}\ and\ \citenamefont
  {Ye}(2003)}]{Cundiff2003}%
  \BibitemOpen
  \bibfield  {author} {\bibinfo {author} {\bibfnamefont {S.~T.}\ \bibnamefont
  {Cundiff}}\ and\ \bibinfo {author} {\bibfnamefont {J.}~\bibnamefont {Ye}},\
  }\href@noop {} {\bibfield  {journal} {\bibinfo  {journal} {Rev. of Mod.
  Phys.}\ }\textbf {\bibinfo {volume} {75}},\ \bibinfo {pages} {325} (\bibinfo
  {year} {2003})}\BibitemShut {NoStop}%
\bibitem [{\citenamefont {Udem}\ \emph {et~al.}(1999)\citenamefont {Udem},
  \citenamefont {Reichert}, \citenamefont {Holzwarth},\ and\ \citenamefont
  {H{\"a}nsch}}]{Udem1999}%
  \BibitemOpen
  \bibfield  {author} {\bibinfo {author} {\bibfnamefont {T.}~\bibnamefont
  {Udem}}, \bibinfo {author} {\bibfnamefont {J.}~\bibnamefont {Reichert}},
  \bibinfo {author} {\bibfnamefont {R.}~\bibnamefont {Holzwarth}}, \ and\
  \bibinfo {author} {\bibfnamefont {T.~W.}\ \bibnamefont {H{\"a}nsch}},\
  }\href@noop {} {\bibfield  {journal} {\bibinfo  {journal} {Phys. Rev. Lett.}\
  }\textbf {\bibinfo {volume} {82}},\ \bibinfo {pages} {3568} (\bibinfo {year}
  {1999})}\BibitemShut {NoStop}%
\bibitem [{\citenamefont {Coddington}\ \emph {et~al.}(2008)\citenamefont
  {Coddington}, \citenamefont {Swann},\ and\ \citenamefont
  {Newbury}}]{Coddington2008}%
  \BibitemOpen
  \bibfield  {author} {\bibinfo {author} {\bibfnamefont {I.}~\bibnamefont
  {Coddington}}, \bibinfo {author} {\bibfnamefont {W.~C.}\ \bibnamefont
  {Swann}}, \ and\ \bibinfo {author} {\bibfnamefont {N.~R.}\ \bibnamefont
  {Newbury}},\ }\href@noop {} {\bibfield  {journal} {\bibinfo  {journal} {Phys.
  Rev. Lett.}\ }\textbf {\bibinfo {volume} {100}},\ \bibinfo {pages} {13902}
  (\bibinfo {year} {2008})}\BibitemShut {NoStop}%
\bibitem [{\citenamefont {Diddams}\ \emph {et~al.}(2001)\citenamefont
  {Diddams}, \citenamefont {Udem}, \citenamefont {Bergquist}, \citenamefont
  {Curtis}, \citenamefont {Drullinger}, \citenamefont {Hollberg}, \citenamefont
  {Itano}, \citenamefont {Lee}, \citenamefont {Oates}, \citenamefont {Vogel},\
  and\ \citenamefont {Wineland}}]{Diddams2001}%
  \BibitemOpen
  \bibfield  {author} {\bibinfo {author} {\bibfnamefont {S.~A.}\ \bibnamefont
  {Diddams}}, \bibinfo {author} {\bibfnamefont {T.}~\bibnamefont {Udem}},
  \bibinfo {author} {\bibfnamefont {J.~C.}\ \bibnamefont {Bergquist}}, \bibinfo
  {author} {\bibfnamefont {E.~A.}\ \bibnamefont {Curtis}}, \bibinfo {author}
  {\bibfnamefont {R.~E.}\ \bibnamefont {Drullinger}}, \bibinfo {author}
  {\bibfnamefont {L.}~\bibnamefont {Hollberg}}, \bibinfo {author}
  {\bibfnamefont {W.~M.}\ \bibnamefont {Itano}}, \bibinfo {author}
  {\bibfnamefont {W.~D.}\ \bibnamefont {Lee}}, \bibinfo {author} {\bibfnamefont
  {C.~W.}\ \bibnamefont {Oates}}, \bibinfo {author} {\bibfnamefont {K.~R.}\
  \bibnamefont {Vogel}}, \ and\ \bibinfo {author} {\bibfnamefont {D.~J.}\
  \bibnamefont {Wineland}},\ }\href@noop {} {\bibfield  {journal} {\bibinfo
  {journal} {Science}\ }\textbf {\bibinfo {volume} {293}},\ \bibinfo {pages}
  {825} (\bibinfo {year} {2001})}\BibitemShut {NoStop}%
\bibitem [{\citenamefont {Cundiff}\ and\ \citenamefont
  {Weiner}(2010)}]{Cundiff2010}%
  \BibitemOpen
  \bibfield  {author} {\bibinfo {author} {\bibfnamefont {S.~T.}\ \bibnamefont
  {Cundiff}}\ and\ \bibinfo {author} {\bibfnamefont {A.~M.}\ \bibnamefont
  {Weiner}},\ }\href@noop {} {\bibfield  {journal} {\bibinfo  {journal} {Nat.
  Phot.}\ }\textbf {\bibinfo {volume} {4}},\ \bibinfo {pages} {760} (\bibinfo
  {year} {2010})}\BibitemShut {NoStop}%
\bibitem [{\citenamefont {Sutter}\ \emph {et~al.}(1999)\citenamefont {Sutter},
  \citenamefont {Steinmeyer}, \citenamefont {Gallmann}, \citenamefont
  {Matuschek}, \citenamefont {Morier-Genoud}, \citenamefont {Keller},
  \citenamefont {Scheuer}, \citenamefont {Angelow},\ and\ \citenamefont
  {Tschudi}}]{Sutter1999}%
  \BibitemOpen
  \bibfield  {author} {\bibinfo {author} {\bibfnamefont {D.~H.}\ \bibnamefont
  {Sutter}}, \bibinfo {author} {\bibfnamefont {G.}~\bibnamefont {Steinmeyer}},
  \bibinfo {author} {\bibfnamefont {L.}~\bibnamefont {Gallmann}}, \bibinfo
  {author} {\bibfnamefont {N.}~\bibnamefont {Matuschek}}, \bibinfo {author}
  {\bibfnamefont {F.}~\bibnamefont {Morier-Genoud}}, \bibinfo {author}
  {\bibfnamefont {U.}~\bibnamefont {Keller}}, \bibinfo {author} {\bibfnamefont
  {V.}~\bibnamefont {Scheuer}}, \bibinfo {author} {\bibfnamefont
  {G.}~\bibnamefont {Angelow}}, \ and\ \bibinfo {author} {\bibfnamefont
  {T.}~\bibnamefont {Tschudi}},\ }\href@noop {} {\bibfield  {journal} {\bibinfo
   {journal} {Opt. Lett.}\ }\textbf {\bibinfo {volume} {24}},\ \bibinfo {pages}
  {631} (\bibinfo {year} {1999})}\BibitemShut {NoStop}%
\bibitem [{\citenamefont {Fermann}\ and\ \citenamefont
  {Hartl}(2013)}]{Fermann2013}%
  \BibitemOpen
  \bibfield  {author} {\bibinfo {author} {\bibfnamefont {M.~E.}\ \bibnamefont
  {Fermann}}\ and\ \bibinfo {author} {\bibfnamefont {I.}~\bibnamefont
  {Hartl}},\ }\href@noop {} {\bibfield  {journal} {\bibinfo  {journal} {Nat.
  Phot.}\ }\textbf {\bibinfo {volume} {7}},\ \bibinfo {pages} {868} (\bibinfo
  {year} {2013})}\BibitemShut {NoStop}%
\bibitem [{\citenamefont {Herr}\ \emph {et~al.}(2012)\citenamefont {Herr},
  \citenamefont {Hartinger}, \citenamefont {Riemensberger}, \citenamefont
  {Wang}, \citenamefont {Gavartin}, \citenamefont {Holzwarth}, \citenamefont
  {Gorodetsky},\ and\ \citenamefont {Kippenberg}}]{Herr2012}%
  \BibitemOpen
  \bibfield  {author} {\bibinfo {author} {\bibfnamefont {T.}~\bibnamefont
  {Herr}}, \bibinfo {author} {\bibfnamefont {K.}~\bibnamefont {Hartinger}},
  \bibinfo {author} {\bibfnamefont {J.}~\bibnamefont {Riemensberger}}, \bibinfo
  {author} {\bibfnamefont {C.~Y.}\ \bibnamefont {Wang}}, \bibinfo {author}
  {\bibfnamefont {E.}~\bibnamefont {Gavartin}}, \bibinfo {author}
  {\bibfnamefont {R.}~\bibnamefont {Holzwarth}}, \bibinfo {author}
  {\bibfnamefont {M.~L.}\ \bibnamefont {Gorodetsky}}, \ and\ \bibinfo {author}
  {\bibfnamefont {T.~J.}\ \bibnamefont {Kippenberg}},\ }\href@noop {}
  {\bibfield  {journal} {\bibinfo  {journal} {Nat. Phot.}\ }\textbf {\bibinfo
  {volume} {6}},\ \bibinfo {pages} {480} (\bibinfo {year} {2012})}\BibitemShut
  {NoStop}%
\bibitem [{\citenamefont {Moskalenko}\ \emph {et~al.}(2017)\citenamefont
  {Moskalenko}, \citenamefont {Koelemeij}, \citenamefont {Williams},\ and\
  \citenamefont {Bente}}]{Moskalenko2017}%
  \BibitemOpen
  \bibfield  {author} {\bibinfo {author} {\bibfnamefont {V.}~\bibnamefont
  {Moskalenko}}, \bibinfo {author} {\bibfnamefont {J.}~\bibnamefont
  {Koelemeij}}, \bibinfo {author} {\bibfnamefont {K.}~\bibnamefont {Williams}},
  \ and\ \bibinfo {author} {\bibfnamefont {E.}~\bibnamefont {Bente}},\
  }\href@noop {} {\bibfield  {journal} {\bibinfo  {journal} {Opt. Letters}\
  }\textbf {\bibinfo {volume} {42}},\ \bibinfo {pages} {1428} (\bibinfo {year}
  {2017})}\BibitemShut {NoStop}%
\bibitem [{\citenamefont {Rosales}\ \emph {et~al.}(2011)\citenamefont
  {Rosales}, \citenamefont {Merghem}, \citenamefont {Martinez}, \citenamefont
  {Akrout}, \citenamefont {Tourrenc}, \citenamefont {Accard}, \citenamefont
  {Lelarge},\ and\ \citenamefont {Ramdane}}]{Rosales2011}%
  \BibitemOpen
  \bibfield  {author} {\bibinfo {author} {\bibfnamefont {R.}~\bibnamefont
  {Rosales}}, \bibinfo {author} {\bibfnamefont {K.}~\bibnamefont {Merghem}},
  \bibinfo {author} {\bibfnamefont {A.}~\bibnamefont {Martinez}}, \bibinfo
  {author} {\bibfnamefont {A.}~\bibnamefont {Akrout}}, \bibinfo {author}
  {\bibfnamefont {J.-P.}\ \bibnamefont {Tourrenc}}, \bibinfo {author}
  {\bibfnamefont {A.}~\bibnamefont {Accard}}, \bibinfo {author} {\bibfnamefont
  {F.}~\bibnamefont {Lelarge}}, \ and\ \bibinfo {author} {\bibfnamefont
  {A.}~\bibnamefont {Ramdane}},\ }\href@noop {} {\bibfield  {journal} {\bibinfo
   {journal} {IEEE J. Sel. Top. Quantum Electron.}\ }\textbf {\bibinfo {volume}
  {17}},\ \bibinfo {pages} {1292} (\bibinfo {year} {2011})}\BibitemShut
  {NoStop}%
\bibitem [{\citenamefont {Delfyett}\ \emph {et~al.}(1992)\citenamefont
  {Delfyett}, \citenamefont {Florez}, \citenamefont {Stoffel}, \citenamefont
  {Gmitter}, \citenamefont {Andreadakis}, \citenamefont {Silberberg},\ and\
  \citenamefont {Heritage}}]{Delfyett1992}%
  \BibitemOpen
  \bibfield  {author} {\bibinfo {author} {\bibfnamefont {P.~J.}\ \bibnamefont
  {Delfyett}}, \bibinfo {author} {\bibfnamefont {L.~T.}\ \bibnamefont
  {Florez}}, \bibinfo {author} {\bibfnamefont {N.}~\bibnamefont {Stoffel}},
  \bibinfo {author} {\bibfnamefont {T.}~\bibnamefont {Gmitter}}, \bibinfo
  {author} {\bibfnamefont {N.~C.}\ \bibnamefont {Andreadakis}}, \bibinfo
  {author} {\bibfnamefont {Y.}~\bibnamefont {Silberberg}}, \ and\ \bibinfo
  {author} {\bibfnamefont {J.~P.}\ \bibnamefont {Heritage}},\ }\href@noop {}
  {\bibfield  {journal} {\bibinfo  {journal} {IEEE J. Quantum Electron.}\
  }\textbf {\bibinfo {volume} {28}},\ \bibinfo {pages} {2203} (\bibinfo {year}
  {1992})}\BibitemShut {NoStop}%
\bibitem [{\citenamefont {Tiemeijer}\ \emph {et~al.}(1989)\citenamefont
  {Tiemeijer}, \citenamefont {Kuindersma}, \citenamefont {Thijs},\ and\
  \citenamefont {Rikken}}]{Tiemeijer1989}%
  \BibitemOpen
  \bibfield  {author} {\bibinfo {author} {\bibfnamefont {L.~F.}\ \bibnamefont
  {Tiemeijer}}, \bibinfo {author} {\bibfnamefont {P.~I.}\ \bibnamefont
  {Kuindersma}}, \bibinfo {author} {\bibfnamefont {P.~J.~A.}\ \bibnamefont
  {Thijs}}, \ and\ \bibinfo {author} {\bibfnamefont {G.~L.~J.}\ \bibnamefont
  {Rikken}},\ }\href@noop {} {\bibfield  {journal} {\bibinfo  {journal} {IEEE
  J. Quantum Electron.}\ }\textbf {\bibinfo {volume} {25}},\ \bibinfo {pages}
  {1385} (\bibinfo {year} {1989})}\BibitemShut {NoStop}%
\bibitem [{\citenamefont {Gioannini}\ \emph {et~al.}(2015)\citenamefont
  {Gioannini}, \citenamefont {Bardella},\ and\ \citenamefont
  {Montrosset}}]{Gioannini2015}%
  \BibitemOpen
  \bibfield  {author} {\bibinfo {author} {\bibfnamefont {M.}~\bibnamefont
  {Gioannini}}, \bibinfo {author} {\bibfnamefont {P.}~\bibnamefont {Bardella}},
  \ and\ \bibinfo {author} {\bibfnamefont {I.}~\bibnamefont {Montrosset}},\
  }\href@noop {} {\bibfield  {journal} {\bibinfo  {journal} {IEEE Sel. Topics
  Quantum Electron.}\ }\textbf {\bibinfo {volume} {21}},\ \bibinfo {pages}
  {1900811} (\bibinfo {year} {2015})}\BibitemShut {NoStop}%
\bibitem [{\citenamefont {Rosales}\ \emph
  {et~al.}(2012{\natexlab{a}})\citenamefont {Rosales}, \citenamefont {Merghem},
  \citenamefont {Calo}, \citenamefont {Bouwmans}, \citenamefont {Krestnikov},
  \citenamefont {Martinez},\ and\ \citenamefont {Ramdane}}]{Rosales2012}%
  \BibitemOpen
  \bibfield  {author} {\bibinfo {author} {\bibfnamefont {R.}~\bibnamefont
  {Rosales}}, \bibinfo {author} {\bibfnamefont {K.}~\bibnamefont {Merghem}},
  \bibinfo {author} {\bibfnamefont {C.}~\bibnamefont {Calo}}, \bibinfo {author}
  {\bibfnamefont {G.}~\bibnamefont {Bouwmans}}, \bibinfo {author}
  {\bibfnamefont {I.}~\bibnamefont {Krestnikov}}, \bibinfo {author}
  {\bibfnamefont {A.}~\bibnamefont {Martinez}}, \ and\ \bibinfo {author}
  {\bibfnamefont {A.}~\bibnamefont {Ramdane}},\ }\href@noop {} {\bibfield
  {journal} {\bibinfo  {journal} {App. Phys. Lett.}\ }\textbf {\bibinfo
  {volume} {101}},\ \bibinfo {pages} {221113} (\bibinfo {year}
  {2012}{\natexlab{a}})}\BibitemShut {NoStop}%
\bibitem [{\citenamefont {Rosales}\ \emph
  {et~al.}(2012{\natexlab{b}})\citenamefont {Rosales}, \citenamefont {Murdoch},
  \citenamefont {Watts}, \citenamefont {Merghem}, \citenamefont {Martinez},
  \citenamefont {Lelarge}, \citenamefont {Accard}, \citenamefont {Barry},\ and\
  \citenamefont {Ramdane}}]{Rosales2012-2}%
  \BibitemOpen
  \bibfield  {author} {\bibinfo {author} {\bibfnamefont {R.}~\bibnamefont
  {Rosales}}, \bibinfo {author} {\bibfnamefont {S.~G.}\ \bibnamefont
  {Murdoch}}, \bibinfo {author} {\bibfnamefont {R.}~\bibnamefont {Watts}},
  \bibinfo {author} {\bibfnamefont {K.}~\bibnamefont {Merghem}}, \bibinfo
  {author} {\bibfnamefont {A.}~\bibnamefont {Martinez}}, \bibinfo {author}
  {\bibfnamefont {F.}~\bibnamefont {Lelarge}}, \bibinfo {author} {\bibfnamefont
  {A.}~\bibnamefont {Accard}}, \bibinfo {author} {\bibfnamefont {L.~P.}\
  \bibnamefont {Barry}}, \ and\ \bibinfo {author} {\bibfnamefont
  {A.}~\bibnamefont {Ramdane}},\ }\href@noop {} {\bibfield  {journal} {\bibinfo
   {journal} {Optics Express}\ }\textbf {\bibinfo {volume} {20}},\ \bibinfo
  {pages} {8649} (\bibinfo {year} {2012}{\natexlab{b}})}\BibitemShut {NoStop}%
\bibitem [{\citenamefont {Sato}(2003)}]{Sato2003}%
  \BibitemOpen
  \bibfield  {author} {\bibinfo {author} {\bibfnamefont {K.}~\bibnamefont
  {Sato}},\ }\href@noop {} {\bibfield  {journal} {\bibinfo  {journal} {IEEE J.
  Sel. Top. Quantum Electron.}\ }\textbf {\bibinfo {volume} {9}},\ \bibinfo
  {pages} {1288} (\bibinfo {year} {2003})}\BibitemShut {NoStop}%
\bibitem [{\citenamefont {Cal{\`o}}\ \emph {et~al.}(2015)\citenamefont
  {Cal{\`o}}, \citenamefont {Vujicic}, \citenamefont {Watts}, \citenamefont
  {Browning}, \citenamefont {Merghem}, \citenamefont {Panapakkam},
  \citenamefont {Lelarge}, \citenamefont {Martinez}, \citenamefont {Benkelfat},
  \citenamefont {Ramdane},\ and\ \citenamefont {Barry}}]{Calo2015}%
  \BibitemOpen
  \bibfield  {author} {\bibinfo {author} {\bibfnamefont {C.}~\bibnamefont
  {Cal{\`o}}}, \bibinfo {author} {\bibfnamefont {V.}~\bibnamefont {Vujicic}},
  \bibinfo {author} {\bibfnamefont {R.}~\bibnamefont {Watts}}, \bibinfo
  {author} {\bibfnamefont {C.}~\bibnamefont {Browning}}, \bibinfo {author}
  {\bibfnamefont {K.}~\bibnamefont {Merghem}}, \bibinfo {author} {\bibfnamefont
  {V.}~\bibnamefont {Panapakkam}}, \bibinfo {author} {\bibfnamefont
  {F.}~\bibnamefont {Lelarge}}, \bibinfo {author} {\bibfnamefont
  {A.}~\bibnamefont {Martinez}}, \bibinfo {author} {\bibfnamefont {B.-E.}\
  \bibnamefont {Benkelfat}}, \bibinfo {author} {\bibfnamefont {A.}~\bibnamefont
  {Ramdane}}, \ and\ \bibinfo {author} {\bibfnamefont {L.~P.}\ \bibnamefont
  {Barry}},\ }\href@noop {} {\bibfield  {journal} {\bibinfo  {journal} {Opt.
  Express}\ }\textbf {\bibinfo {volume} {23}},\ \bibinfo {pages} {26442}
  (\bibinfo {year} {2015})}\BibitemShut {NoStop}%
\bibitem [{\citenamefont {Homar}\ \emph
  {et~al.}(1996{\natexlab{a}})\citenamefont {Homar}, \citenamefont {Balle},\
  and\ \citenamefont {Miguel}}]{Homar1996}%
  \BibitemOpen
  \bibfield  {author} {\bibinfo {author} {\bibfnamefont {M.}~\bibnamefont
  {Homar}}, \bibinfo {author} {\bibfnamefont {S.}~\bibnamefont {Balle}}, \ and\
  \bibinfo {author} {\bibfnamefont {M.~S.}\ \bibnamefont {Miguel}},\
  }\href@noop {} {\bibfield  {journal} {\bibinfo  {journal} {Optics
  Communications}\ }\textbf {\bibinfo {volume} {131}},\ \bibinfo {pages} {380}
  (\bibinfo {year} {1996}{\natexlab{a}})}\BibitemShut {NoStop}%
\bibitem [{\citenamefont {Arakawa}\ and\ \citenamefont
  {Yariv}(1986)}]{Arakawa1986}%
  \BibitemOpen
  \bibfield  {author} {\bibinfo {author} {\bibfnamefont {Y.}~\bibnamefont
  {Arakawa}}\ and\ \bibinfo {author} {\bibfnamefont {A.}~\bibnamefont
  {Yariv}},\ }\href@noop {} {\bibfield  {journal} {\bibinfo  {journal} {IEEE J.
  Quantum Electron.}\ }\textbf {\bibinfo {volume} {QE22}},\ \bibinfo {pages}
  {1887} (\bibinfo {year} {1986})}\BibitemShut {NoStop}%
\bibitem [{\citenamefont {Kaunga-Nyirenda}\ \emph {et~al.}(2010)\citenamefont
  {Kaunga-Nyirenda}, \citenamefont {Dlubek}, \citenamefont {Phillips},
  \citenamefont {Lim}, \citenamefont {Larkins},\ and\ \citenamefont
  {Sujecki}}]{KN2010}%
  \BibitemOpen
  \bibfield  {author} {\bibinfo {author} {\bibfnamefont {S.~N.}\ \bibnamefont
  {Kaunga-Nyirenda}}, \bibinfo {author} {\bibfnamefont {M.~P.}\ \bibnamefont
  {Dlubek}}, \bibinfo {author} {\bibfnamefont {A.~J.}\ \bibnamefont
  {Phillips}}, \bibinfo {author} {\bibfnamefont {J.~J.}\ \bibnamefont {Lim}},
  \bibinfo {author} {\bibfnamefont {E.~C.}\ \bibnamefont {Larkins}}, \ and\
  \bibinfo {author} {\bibfnamefont {S.}~\bibnamefont {Sujecki}},\ }\href@noop
  {} {\bibfield  {journal} {\bibinfo  {journal} {J. Opt. Soc. Am. B}\ }\textbf
  {\bibinfo {volume} {27}},\ \bibinfo {pages} {168} (\bibinfo {year}
  {2010})}\BibitemShut {NoStop}%
\bibitem [{\citenamefont {McDonald}\ and\ \citenamefont
  {O'Dowd}(1995)}]{McDonald1995}%
  \BibitemOpen
  \bibfield  {author} {\bibinfo {author} {\bibfnamefont {D.}~\bibnamefont
  {McDonald}}\ and\ \bibinfo {author} {\bibfnamefont {R.~F.}\ \bibnamefont
  {O'Dowd}},\ }\href@noop {} {\bibfield  {journal} {\bibinfo  {journal} {IEEE
  J. Quantum Electron.}\ }\textbf {\bibinfo {volume} {31}},\ \bibinfo {pages}
  {1927} (\bibinfo {year} {1995})}\BibitemShut {NoStop}%
\bibitem [{\citenamefont {Jones}\ \emph {et~al.}(1995)\citenamefont {Jones},
  \citenamefont {Zhang}, \citenamefont {Carroll},\ and\ \citenamefont
  {Marcenac}}]{Jones1995}%
  \BibitemOpen
  \bibfield  {author} {\bibinfo {author} {\bibfnamefont {D.~J.}\ \bibnamefont
  {Jones}}, \bibinfo {author} {\bibfnamefont {L.~M.}\ \bibnamefont {Zhang}},
  \bibinfo {author} {\bibfnamefont {J.~E.}\ \bibnamefont {Carroll}}, \ and\
  \bibinfo {author} {\bibfnamefont {D.~D.}\ \bibnamefont {Marcenac}},\
  }\href@noop {} {\bibfield  {journal} {\bibinfo  {journal} {IEEE J. Quantum
  Electron.}\ }\textbf {\bibinfo {volume} {31}},\ \bibinfo {pages} {1051}
  (\bibinfo {year} {1995})}\BibitemShut {NoStop}%
\bibitem [{\citenamefont {Vandermeer}\ and\ \citenamefont
  {Cassidy}(2005)}]{Vandermeer2005}%
  \BibitemOpen
  \bibfield  {author} {\bibinfo {author} {\bibfnamefont {A.~D.}\ \bibnamefont
  {Vandermeer}}\ and\ \bibinfo {author} {\bibfnamefont {D.~T.}\ \bibnamefont
  {Cassidy}},\ }\href@noop {} {\bibfield  {journal} {\bibinfo  {journal} {IEEE
  J. Quantum Electron.}\ }\textbf {\bibinfo {volume} {41}},\ \bibinfo {pages}
  {917} (\bibinfo {year} {2005})}\BibitemShut {NoStop}%
\bibitem [{\citenamefont {Gordon}\ \emph {et~al.}(2008)\citenamefont {Gordon},
  \citenamefont {Wang}, \citenamefont {Diehl}, \citenamefont {K{\"a}rtner},
  \citenamefont {Belyanin}, \citenamefont {Bour}, \citenamefont {Corzine},
  \citenamefont {H{\"o}fler}, \citenamefont {Liu}, \citenamefont {Schneider},
  \citenamefont {Maier}, \citenamefont {Troccoli}, \citenamefont {Faist},\ and\
  \citenamefont {Capasso}}]{Gordon2008}%
  \BibitemOpen
  \bibfield  {author} {\bibinfo {author} {\bibfnamefont {A.}~\bibnamefont
  {Gordon}}, \bibinfo {author} {\bibfnamefont {C.~Y.}\ \bibnamefont {Wang}},
  \bibinfo {author} {\bibfnamefont {L.}~\bibnamefont {Diehl}}, \bibinfo
  {author} {\bibfnamefont {F.~X.}\ \bibnamefont {K{\"a}rtner}}, \bibinfo
  {author} {\bibfnamefont {A.}~\bibnamefont {Belyanin}}, \bibinfo {author}
  {\bibfnamefont {D.}~\bibnamefont {Bour}}, \bibinfo {author} {\bibfnamefont
  {S.}~\bibnamefont {Corzine}}, \bibinfo {author} {\bibfnamefont
  {G.}~\bibnamefont {H{\"o}fler}}, \bibinfo {author} {\bibfnamefont {H.~C.}\
  \bibnamefont {Liu}}, \bibinfo {author} {\bibfnamefont {H.}~\bibnamefont
  {Schneider}}, \bibinfo {author} {\bibfnamefont {T.}~\bibnamefont {Maier}},
  \bibinfo {author} {\bibfnamefont {M.}~\bibnamefont {Troccoli}}, \bibinfo
  {author} {\bibfnamefont {J.}~\bibnamefont {Faist}}, \ and\ \bibinfo {author}
  {\bibfnamefont {F.}~\bibnamefont {Capasso}},\ }\href@noop {} {\bibfield
  {journal} {\bibinfo  {journal} {Phys. Rev. A}\ }\textbf {\bibinfo {volume}
  {77}},\ \bibinfo {pages} {053804} (\bibinfo {year} {2008})}\BibitemShut
  {NoStop}%
\bibitem [{\citenamefont {Lenstra}\ and\ \citenamefont
  {Yousefi}(2014)}]{Lenstra2014}%
  \BibitemOpen
  \bibfield  {author} {\bibinfo {author} {\bibfnamefont {D.}~\bibnamefont
  {Lenstra}}\ and\ \bibinfo {author} {\bibfnamefont {M.}~\bibnamefont
  {Yousefi}},\ }\href@noop {} {\bibfield  {journal} {\bibinfo  {journal} {Opt.
  Express}\ }\textbf {\bibinfo {volume} {22}},\ \bibinfo {pages} {8144}
  (\bibinfo {year} {2014})}\BibitemShut {NoStop}%
\bibitem [{\citenamefont {Chow}\ \emph {et~al.}(2002)\citenamefont {Chow},
  \citenamefont {Schneider}, \citenamefont {Koch}, \citenamefont {Chang},
  \citenamefont {Chrostowski},\ and\ \citenamefont {Chang-Hasnain}}]{Chow2002}%
  \BibitemOpen
  \bibfield  {author} {\bibinfo {author} {\bibfnamefont {W.~W.}\ \bibnamefont
  {Chow}}, \bibinfo {author} {\bibfnamefont {H.~C.}\ \bibnamefont {Schneider}},
  \bibinfo {author} {\bibfnamefont {S.~W.}\ \bibnamefont {Koch}}, \bibinfo
  {author} {\bibfnamefont {C.-H.}\ \bibnamefont {Chang}}, \bibinfo {author}
  {\bibfnamefont {L.}~\bibnamefont {Chrostowski}}, \ and\ \bibinfo {author}
  {\bibfnamefont {C.~J.}\ \bibnamefont {Chang-Hasnain}},\ }\href@noop {}
  {\bibfield  {journal} {\bibinfo  {journal} {IEEE J. Quantum Electron.}\
  }\textbf {\bibinfo {volume} {38}},\ \bibinfo {pages} {402} (\bibinfo {year}
  {2002})}\BibitemShut {NoStop}%
\bibitem [{\citenamefont {Chuang}(2009)}]{Chuang2009}%
  \BibitemOpen
  \bibfield  {author} {\bibinfo {author} {\bibfnamefont {S.~L.}\ \bibnamefont
  {Chuang}},\ }\href@noop {} {\emph {\bibinfo {title} {Physics of Photonic
  Devices}}},\ \bibinfo {edition} {2nd}\ ed.\ (\bibinfo  {publisher} {John
  Wiley and Sons, Inc.},\ \bibinfo {address} {Hoboken, NJ},\ \bibinfo {year}
  {2009})\BibitemShut {NoStop}%
\bibitem [{\citenamefont {Chow}\ \emph {et~al.}(1994)\citenamefont {Chow},
  \citenamefont {Koch},\ and\ \citenamefont {III}}]{ChowKochSargent1994}%
  \BibitemOpen
  \bibfield  {author} {\bibinfo {author} {\bibfnamefont {W.~W.}\ \bibnamefont
  {Chow}}, \bibinfo {author} {\bibfnamefont {S.~W.}\ \bibnamefont {Koch}}, \
  and\ \bibinfo {author} {\bibfnamefont {M.~S.}\ \bibnamefont {III}},\
  }\href@noop {} {\emph {\bibinfo {title} {Semiconductor-Laser Physics}}}\
  (\bibinfo  {publisher} {Springer-Verlag},\ \bibinfo {year}
  {1994})\BibitemShut {NoStop}%
\bibitem [{\citenamefont {Javaloyes}\ and\ \citenamefont
  {Balle}(2009)}]{Javaloyes2009}%
  \BibitemOpen
  \bibfield  {author} {\bibinfo {author} {\bibfnamefont {J.}~\bibnamefont
  {Javaloyes}}\ and\ \bibinfo {author} {\bibfnamefont {S.}~\bibnamefont
  {Balle}},\ }\href@noop {} {\bibfield  {journal} {\bibinfo  {journal} {IEEE J.
  Quantum Electron.}\ }\textbf {\bibinfo {volume} {45}},\ \bibinfo {pages}
  {431} (\bibinfo {year} {2009})}\BibitemShut {NoStop}%
\bibitem [{\citenamefont {Homar}\ \emph
  {et~al.}(1996{\natexlab{b}})\citenamefont {Homar}, \citenamefont {Moloney},\
  and\ \citenamefont {Miguel}}]{Homar1996-2}%
  \BibitemOpen
  \bibfield  {author} {\bibinfo {author} {\bibfnamefont {M.}~\bibnamefont
  {Homar}}, \bibinfo {author} {\bibfnamefont {J.~V.}\ \bibnamefont {Moloney}},
  \ and\ \bibinfo {author} {\bibfnamefont {M.~S.}\ \bibnamefont {Miguel}},\
  }\href@noop {} {\bibfield  {journal} {\bibinfo  {journal} {IEEE J. Quantum
  Electron.}\ }\textbf {\bibinfo {volume} {32}},\ \bibinfo {pages} {553}
  (\bibinfo {year} {1996}{\natexlab{b}})}\BibitemShut {NoStop}%
\bibitem [{\citenamefont {Rossetti}\ \emph {et~al.}(2011)\citenamefont
  {Rossetti}, \citenamefont {Bardella},\ and\ \citenamefont
  {Montrosset}}]{Rossetti2011}%
  \BibitemOpen
  \bibfield  {author} {\bibinfo {author} {\bibfnamefont {M.}~\bibnamefont
  {Rossetti}}, \bibinfo {author} {\bibfnamefont {P.}~\bibnamefont {Bardella}},
  \ and\ \bibinfo {author} {\bibfnamefont {I.}~\bibnamefont {Montrosset}},\
  }\href@noop {} {\bibfield  {journal} {\bibinfo  {journal} {IEEE J. Quantum
  Electron.}\ }\textbf {\bibinfo {volume} {47}},\ \bibinfo {pages} {139}
  (\bibinfo {year} {2011})}\BibitemShut {NoStop}%
\bibitem [{\citenamefont {Marshall}\ \emph {et~al.}(2000)\citenamefont
  {Marshall}, \citenamefont {Miller},\ and\ \citenamefont
  {Button}}]{Marshall2000}%
  \BibitemOpen
  \bibfield  {author} {\bibinfo {author} {\bibfnamefont {D.}~\bibnamefont
  {Marshall}}, \bibinfo {author} {\bibfnamefont {A.}~\bibnamefont {Miller}}, \
  and\ \bibinfo {author} {\bibfnamefont {C.~C.}\ \bibnamefont {Button}},\
  }\href@noop {} {\bibfield  {journal} {\bibinfo  {journal} {IEEE J. Quantum
  Electron.}\ }\textbf {\bibinfo {volume} {36}},\ \bibinfo {pages} {1013}
  (\bibinfo {year} {2000})}\BibitemShut {NoStop}%
\bibitem [{\citenamefont {Su}(1988)}]{Su1988}%
  \BibitemOpen
  \bibfield  {author} {\bibinfo {author} {\bibfnamefont {C.~B.}\ \bibnamefont
  {Su}},\ }\href@noop {} {\bibfield  {journal} {\bibinfo  {journal} {IEEE
  Electron. Lett.}\ }\textbf {\bibinfo {volume} {24}},\ \bibinfo {pages} {370}
  (\bibinfo {year} {1988})}\BibitemShut {NoStop}%
\end{thebibliography}%

\newpage
\begin{table}[!h]
\begin{center}
\begin{tabular}{|c|c|c|}
\hline
Parameter & Description & Value\\ \hline
$L$ & Length of device & 500 $\mu$m \\ \hline
$W$ & Width of waveguide & 4 $\mu$m \\ \hline
$h_{sch}$ & Height of SCH layer & 50 nm \\ \hline
$h_{qw}$ & Height of quantum well & 5 nm \\ \hline
$n_0$ & Group refractive index & 3.5 \\ \hline
$n_{qw}$ & Number of quantum wells & 2 \\ \hline
$\alpha$ & Intrinsic waveguide loss & 5 cm$^{-1}$ \\ \hline
$\Gamma_{xy}$ & Optical confinement factor & 0.01 \\ \hline
$\alpha_S$ & Two-photon absorption & 2750 W$^{-1}$m$^{-1}$\\ \hline
$\beta_S$ & Kerr coefficient &  430 W$^{-1}$m$^{-1}$\\ \hline
$k''$ & Dispersion coefficient &  1.25 ps$^2$/m\\ \hline
$\hbar \omega_0$ & Central transition energy & 800 meV\\ \hline
$|\uv{e}\cdot\v{p}|^2 $ & Momentum matrix element & 21 meV $\times m_0/6$ \cite{Chuang2009}\\ \hline
$\Gamma$ & Homogenous half linewidth & 10 meV/$\hbar$ \\ \hline
$m^*_{e,h, sch}$ & Effective mass of electrons, holes in the SCH layer & $0.07 m_0$, $0.55 m_0$ \\ \hline
$m^*_{e,h, qw}$ & Effective mass of electrons, holes, in the InGaAsP QW & $0.067 m_0$, $0.45 m_0$ \\ \hline
$\tau_c^{e, h,qw}$ & electron, hole capture time  & $1$, 10 ps \\ \hline
$\delta E_c$ & Conduction band quantum well barrier & 50 meV\\ \hline
$\delta E_v$ & Valence band quantum well barrier &  75 meV\\ \hline
$\beta_{sp}$ & Spontaneous emission coupling factor& $1\times 10^{-4}$ \\ \hline
$\tau_{sp}$ & Spontaneous emission lifetime & 1 ns \\ \hline
$D$ & Ambipolar diffusion coefficient & 7.2 cm$^2$/s \cite{Marshall2000}  \\ \hline
\end{tabular}
\end{center}
\caption{Simulation parameters for QW traveling wave model for the InGaAsP system.}
\label{material_param}
\end{table}

\newpage

\end{document}